# Lie Generator Networks Extract EIS-Grade Battery Diagnostics from Pulse Relaxation Data

Shafayeth Jamil, and Rehan Kapadia

*Department of Electrical and Computer Engineering*
*University of Southern California*
sjamil@usc.edu, rkapadia@usc.edu

## Abstract

Electrochemical impedance spectroscopy (EIS) is the most informative diagnostic for lithium-ion batteries: its frequency-resolved spectra decompose cell behavior into distinct electrochemical processes, revealing mechanism-specific degradation invisible to voltage and resistance measurements. Yet EIS requires dedicated hardware and minutes-long acquisitions incompatible with field deployment. Here we show that Lie Generator Networks (LGN), a structure-preserving identification framework, extract electrochemical time constants from 60 seconds of post-pulse voltage relaxation, data that battery management systems already collect, that encode the same diagnostic and prognostic information as impedance spectra. LGN learns the generator matrix of the relaxation dynamics with stability guaranteed by architecture, yielding time constants precise enough to resolve electrochemical variation that conventional curve fitting cannot detect from identical data. Across five datasets totaling over 850 cells, four institutions, and multiple chemistries LGN tracks degradation with near-perfect rank correlation ($|\rho_s| = 0.999$), enables cross-validated reconstruction of full Nyquist spectra at 2% median error across 227 cells, predicts which capacity-matched cells fail first from three early diagnostics, and recovers Arrhenius activation energies with zero physics priors without retraining or cell-specific tuning. LGN requires no training data, no impedance hardware, and no chemistry-specific calibration, converting any existing relaxation pulse into an impedance-grade diagnostic. This enables real-time health monitoring, rapid second-life grading, production-line quality control, and physics-informed prognosis from minutes of measurement.

## Introduction

Lithium-ion batteries underpin the electrification of transport and the integration of renewable energy into the grid, with global battery demand surpassing 1 TWh in 2024, a milestone where a single week's demand now exceeds an entire year's production from a decade earlier [1]. As these fleets age, the ability to assess individual cell health rapidly and accurately becomes critical, not only for safe operation during first life, but increasingly for the sorting and grading of retired cells entering second-life stationary storage, where the absence of historical usage data demands diagnostic methods that can characterize a cell's electrochemical state from a single, brief measurement [2]. Today's battery management systems operate within severe constraints requiring onboard sensors to record voltage, current, and temperature, although the diagnostic inference is limited to quantities derivable from these signals such as open-circuit voltage, coulomb-counted capacity, and DC internal resistance [3]. These aggregate metrics track bulk degradation trends but are fundamentally insensitive to the mechanism-specific processes such as interfacial kinetics, solid-electrolyte interphase growth, active surface area loss that govern how a cell will degrade in the future and distinguish cells with identical present capacity but divergent remaining lifetimes.

EIS resolves the problem by probing a cell with sinusoidal perturbations across frequencies spanning millihertz to megahertz. It yields impedance spectra whose deconvolution through distribution-of-relaxation-times (DRT) analysis separates the contributions of charge transfer, solid-electrolyte interphase transport, and solid-state diffusion into distinct spectral peaks; a fingerprint of the cell's internal electrochemical state [4], [5],[6]. Machine learning applied to large EIS datasets has confirmed that these spectral features encode degradation trajectories with high fidelity [7]. Yet EIS remains a laboratory technique since each measurement requires dedicated instrumentation, controlled environmental conditions, and 10–60 minutes of acquisition time per spectrum. These constraints are incompatible with the throughput demands of fleet-scale monitoring or the rapid sorting of thousands of retired cells entering second-life markets [8].

Efforts to bridge this gap between electrochemical richness and field-deployable speed have followed three largely independent paths. The first relies on machine learning applied to cycling data: statistical features extracted from early charge–discharge curves can predict lifetime with high accuracy, but require tens to hundreds of full cycles as input and yield scalar predictions with no mechanistic decomposition [9], [10]. The second pursues pulse-to-impedance reconstruction, recovering partial impedance spectra from short current pulses without a potentiostat [11], [12]. These methods reduce measurement time by orders of magnitude but treat the reconstructed spectrum as an end product rather than decomposing it into the physical time constants that underlie degradation. The third fits equivalent circuit model parameters, resistances and capacitances, directly to pulse or impedance data, but the resulting nonlinear optimization landscape is ill-conditioned as multiple parameter sets can fit the same waveform equally well, and small perturbations in data produce large swings in extracted values [13].

What remains missing is a method that extracts physically interpretable electrochemical time constants — not statistical proxies, not reconstructed spectra, not scalar health indices — directly from seconds of pulse data, with mathematical guarantees on stability, without requiring paired impedance measurements for training, and without retraining when the battery chemistry changes. Here we apply Lie Generator Networks (LGN) [14], a structure-preserving linear system identification framework, to battery diagnostics and show that a single method meets all of these requirements simultaneously. LGN learns the generator matrix of a linear dynamical system from short pulse relaxation data, whose eigenvalues encode the electrochemical time scales of charge transfer, solid-electrolyte interphase dynamics, and solid-state diffusion. Stability is guaranteed by construction — all eigenvalues are confined to the negative real half-plane by architecture, not regularization — eliminating the ill-conditioning that plagues conventional curve fitting. We validate the method across five independent datasets totaling more than 850 cells from four institutions, spanning NMC and NCA cathodes with graphite and silicon-containing anodes. In every dataset, LGN extracts more diagnostic information from 60 seconds of voltage relaxation, using data that battery management systems already collect during normal operation, than conventional methods obtain from extensive impedance sweeps or charge–discharge cycles, and does so without training data, without impedance measurements, and without cell-specific tuning. We demonstrate four applications: longitudinal degradation tracking that matches or exceeds EIS fidelity, early-life prognosis from time-constant trajectories that encode information orthogonal to capacity, single-pulse manufacturing fingerprinting at 254-cell production scale, and cross-modal Arrhenius and Butler-Volmer kinetics validation confirming that LGN recovers genuine electrochemical processes with zero physics priors.

## Framework

An $n^{\mathrm{th}}$ order equivalent circuit model of a lithium-ion cell comprises $n$ parallel RC branches in series. When the applied current returns to zero after a pulse, the branch voltages evolve as an autonomous linear system:

$$\dot{v}(t) = Av(t), v(t) = e^{At}v(0),$$

where $v(t) \in \mathbb{R}^{\mathrm{n}}$ is the vector of branch voltages, $A \in \mathbb{R}^{\mathrm{nxn}}$ is the system matrix and the measurable terminal overpotential is

$$\eta(t) \equiv V_{\mathrm{term}}(t) - V_{\infty} = \mathbf{1}^{\top}v(t)$$

where $V_{\infty}$ is the post-relaxation equilibrium voltage.

For a diagonal $A$ with entries $-1/\tau_i$, the solution is a sum of exponentials and the impedance transfer function has poles at the same characteristic times:

$$Z(\omega) = R_0 + \sum_{i=1}^{n} \frac{R_i}{1 + j\omega\tau_i}, \tau_i = R_i C_i.$$

Thus, identifying $\{\tau_i\}$ from time-domain pulse relaxation is directly linked to frequency-domain EIS.

Lie Generator Networks (LGN) learn a linear generator $A$ and propagate states via matrix exponentiation rather than numerical integration. For general linear systems, LGN uses a structured parameterization

$$A = S - D,$$

where $S = -S^{\top}$ is skew-symmetric (energy-preserving) and $D \succ 0$ is diagonal (dissipative). Post-pulse relaxation is purely dissipative, so throughout this work we use the minimal relaxation parameterization

$$A = -D, D = \mathrm{diag}\,(d_1, \dots, d_n), d_i > 0.$$

This guarantees stability and dissipation by construction:

$$\frac{d}{dt} \| v(t) \|^2 = 2v^\top \dot{v} = -2v^\top D v \leq 0.$$

The characteristic times are

$$\tau_i = \frac{1}{d_i},$$

with positivity enforced by a softplus map on the learned parameters. Propagation by $e^{At}$ is exact for linear systems and has no step-size dependence or stiffness-induced truncation error. Full architectural and optimization details can be found in the LGN paper [14].

Given a measured post-pulse overpotential η(t) sampled at times $t_k$, LGN recovers the generator by minimizing the reconstruction loss

$$\mathcal{L}(\theta) = \sum_{k=1}^{T} |\eta(t_k) - \mathbf{1}^\top exp(A_\theta t_k) v_0 |^2$$

where $A_\theta = -diag\big(softplus(\theta)\big)$ and $v_0$ are jointly optimized. Gradients propagate through the matrix exponential via automatic differentiation, coupling all eigenvalues during optimization; structurally distinct from unconstrained fitting, where individual basis functions are optimized independently and mode-swapping instabilities arise from the resulting flat directions in the loss landscape (Discussion).

A real battery's relaxation spectrum is distributed rather than discrete as particle-size variation, interphase heterogeneity, and tortuous transport pathways spread each electrochemical process across a continuum of time scales. LGN with n = 3 provides a reduced-order projection of this continuum; fitting n = 2 through 5 to full-length relaxation data confirms three persistent modes corresponding to charge transfer, SEI/interphase transport, and solid-state diffusion, with additional modes acquiring negligible amplitude or collapsing onto existing ones (Figure S1). Training uses warm-starting from the previous aging state, exploiting the physical prior that electrochemical parameters evolve gradually.

## Degradation tracking

We first evaluate LGN on the KIT degradation dataset [15]: 227 NMC/SiO cells aged under diverse protocols with periodic diagnostic checkpoints comprising paired HPPC pulses and full EIS sweeps (0.05–14,700 Hz) at 50% SOC and room temperature. Each cell contributes between 5 and 25 checkpoints spanning beginning-of-life through deep degradation. LGN fits a 3-state model to the voltage relaxation following a single HPPC discharge pulse, extracting three time constants ($\tau_1$, $\tau_2$, $\tau_3$) and their associated resistances ($R_1$, $R_2$, $R_3$) with no per-cell calibration or chemistry-specific tuning. Within each cell, $\tau_1$ and SOH share an almost exact monotonic relationship: the median within-cell Spearman $|\rho_s|$ = 0.999 across 193 cells with ≥5 diagnostic checkpoints (Figure 1a). This is a per-cell rank-order agreement computed independently for each cell, then aggregated. At this level of correlation, $\tau_1$ and SOH are effectively interchangeable as degradation coordinates. Figure 1b places this in context by comparing diagnostic power against acquisition time. LGN $\tau_1$ achieves its $|\rho_s| = 0.999$ from a one-minute pulse — data that battery management systems already collect during normal operation. The best single-frequency EIS feature, cherry-picked per cell from five representative frequencies spanning the full sweep, achieves $|\rho_s| = 0.996$ while requiring ~13 minutes of dedicated frequency-domain excitation and impedance hardware that field-deployed systems lack. Pulse resistance $R_{pulse}$, extracted from the same one-minute measurement, reaches only $|\rho_s| = 0.980$. LGN thus provides EIS-grade tracking accuracy from pulse data alone, without additional hardware or measurement time. The origin of this sensitivity advantage is mechanistic. Decomposing the time constant into its constituents (Figure 1c) shows that from beginning to end of life, $R_1$ increases ~3.5× as interfaces degrade, while $C_1 = \tau_1/R_1$ collapses to ~4% of its initial value — consistent with progressive loss of electrochemically active surface area. The time constant $\tau_1 = R_1C_1$ captures the product of these co-evolving quantities: measuring resistance alone partially masks the capacitance collapse, but the time constant integrates both signals into a single, more sensitive degradation coordinate.

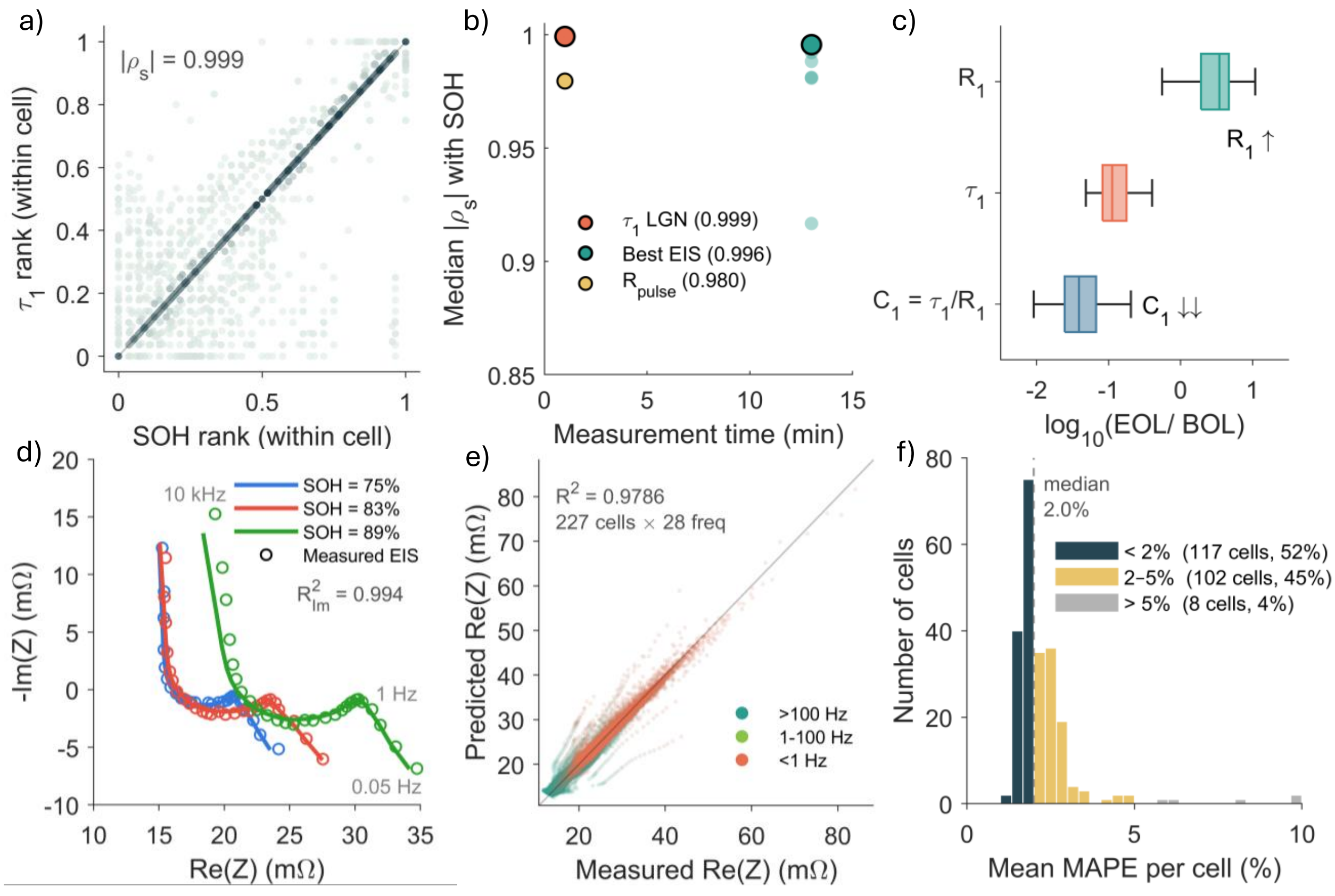

**Figure 1 | Sixty seconds of HPPC voltage relaxation matches or exceeds EIS-grade battery diagnostics across 227 cells** (KIT degradation dataset ; NMC/SiO chemistry, SOC 50%, room temperature). LGN extracts three time constants ($\tau_1$ , $\tau_2$ , $\tau_3$) and their associated resistances from the voltage relaxation following a standard HPPC discharge pulse **(a)** Within-cell percentile-rank agreement between $\tau_1$and SOH across diagnostic checkpoints; median Spearman | $\rho_s$ |= 0.999 across 193 cells with ≥ 5 checkups. **(b)** Pareto comparison of diagnostic accuracy versus measurement time. Each teal marker represents the median $|\rho_s|$ of a single EIS impedance feature (0.05 Hz–14.7 kHz, 29 frequencies, mean acquisition time 12.7 min); "Best EIS" selects the strongest per cell. LGN $\tau_1$ achieves higher correlation in 1/13th the acquisition time with no additional hardware. **(c)** Mechanistic decomposition: EOL/BOL ratios for $R_1$, $\tau_1$, and $C_1 = \tau_1/R_1$ ($R_1$ recovered by fitting amplitudes of the relaxation curve with LGN time constants fixed). $R_1$ grows 3.5× while $C_1$ collapses to 4%, consistent with active surface area loss. **(d)** Nyquist spectra reconstructed from pulse-derived features alone. Open circles: measured EIS; lines: leave-one-cell-out (LOOCV) predictions ($R_{IM}^2$ = 0.994) from a linear map of seven LGN parameters ($R_s$, $\tau_{1-3}$, $R_{1-3}$) to complex impedance. No EIS data from the held-out cell is used during fitting. **(e)** Parity plot of predicted versus measured Re(Z) across all 227 cells and 28 frequencies ($R_{IM}^2$ = 0.979), colored by frequency band. **(f)** Distribution of per-cell mean absolute percentage error (MAPE) from LOOCV impedance reconstruction. Median 2.0%; 96% < 5% MAPE.

Beyond tracking, the seven LGN parameters extracted from each pulse ($R_s$, $\tau_{1-3}$, $R_{1-3}$) encode sufficient information to reconstruct the cell's full complex impedance spectrum. We fit a linear ridge regression from these seven features to complex impedance at each of 28 frequencies using leave-one-cell-out cross-validation (LOOCV), where the model trains on all checkpoints from 226 cells and predicts the held-out cell's spectrum entirely from its pulse-derived features — no EIS data from the test cell enters fitting at any stage. The reconstructed Nyquist spectra track measured EIS across the full frequency range (Figure 1d), capturing the high-frequency inductive response, charge-transfer semicircle, and low-frequency diffusion tail with $R^2_{Im(Z)} = 0.994$. Across all 227 cells and 28 frequencies, predicted versus measured Re(Z) yields $R^2_{Re(Z)} = 0.979$ (Figure 1e), with accuracy maintained from the high-frequency regime through the sub-1 Hz diffusion region where impedance spans nearly an order of magnitude. The per-cell MAPE distribution confirms this is not driven by a few well-behaved cells: the median is 2.0%, with 52% of cells below 2% and 96% below 5% (Figure 1f). That a linear map from seven pulse-derived numbers reconstructs full impedance spectra across 227 cells — without nonlinear fitting, neural networks, or any test-cell EIS data — establishes a direct, physics-grounded bridge between time-domain pulse identification and frequency-domain impedance spectroscopy.

## Early-life prognosis

To test whether pulse-extracted time constants carry prognostic information, not just tracking current health but predicting future degradation, we turn to the TRI Aging Matrix [10]: 374 NCA/graphite+$SiO_x$ 21700 cells cycled under 207 distinct protocols with periodic HPPC diagnostics at 78% SOC and 25 °C. This dataset spans the full spectrum of aging outcomes, from cells that retain >95% capacity throughout testing to cells that degrade below 40% SOH, with 213 of 320 qualifying cells reaching the conventional 80% end-of-life threshold (Figure 2c). LGN fits a 3-state model to each 40-second pulse relaxation, extracting three time constants and their slopes over the early diagnostic history.

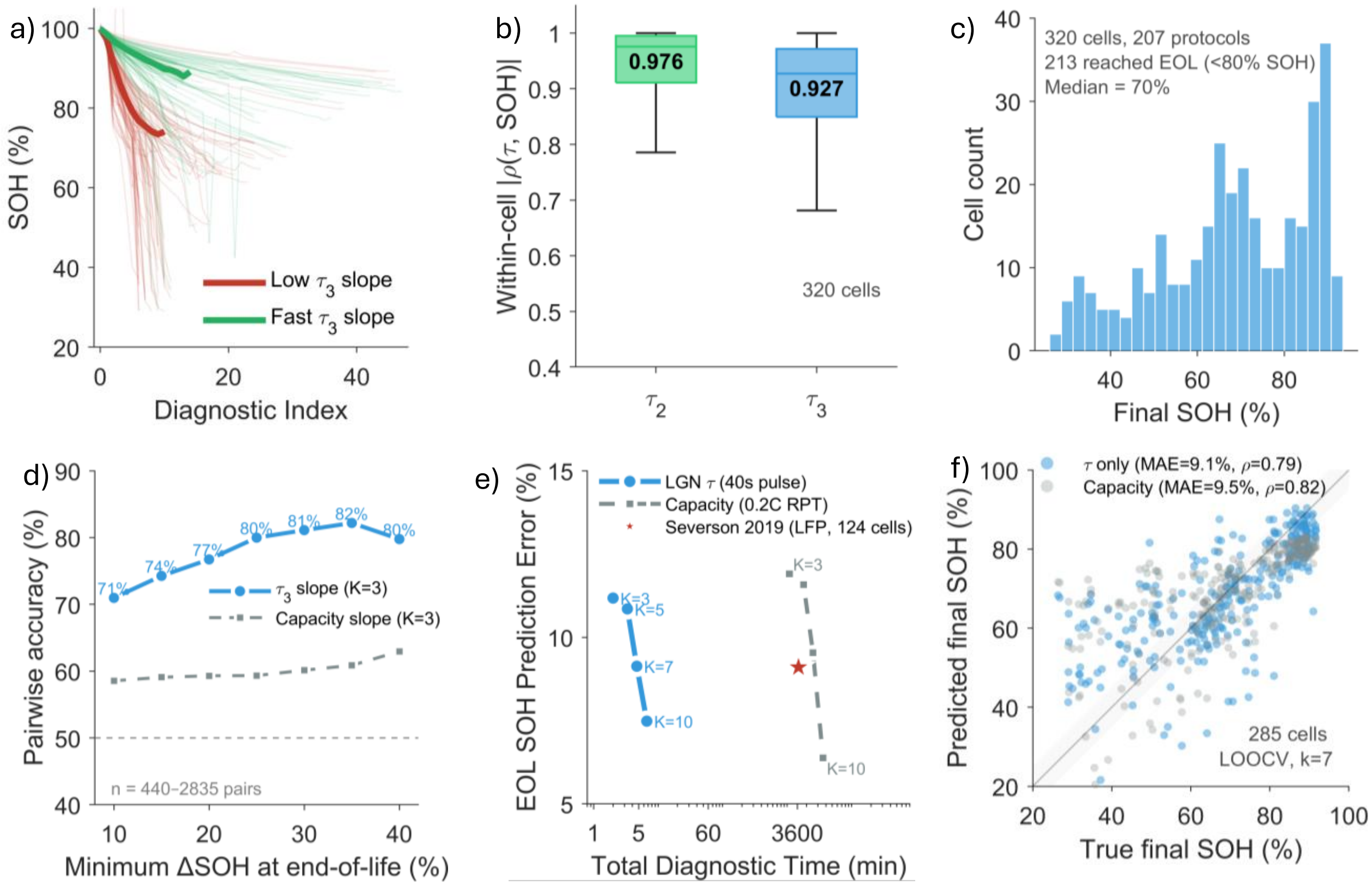

**Figure 2 | Early-life time constants predict end-of-life outcomes across 374 cells and 207 cycling protocols (** TRI Aging Matrix, NCA/graphite+$SiO_x$ 21700 cells, SOC 78%, 25 °C) (a) SOH trajectories grouped by $\tau_3$ slope from the first three diagnostics (K = 3). Cells are matched on initial capacity (±0.15 Ah of median) and split into terciles by early $\tau_3$ drift rate; faint lines show individual cells (n = 93 per group), bold lines show group medians. Low $\tau_3$ slope cells reach 80% EOL within ~15 diagnostics; fast $\tau_3$ slope cells remain above 80% throughout. (b) Within-cell Spearman correlation between time constants and SOH. Median $|\rho_s|$ = 0.976 for $\tau_2$ and 0.927 for $\tau_3$ across 320 cells (≥5 diagnostics each). (c) Distribution of final SOH. 213 of 320 cells reached EOL (<80% SOH); median final SOH = 70%. (d) Pairwise prognostic accuracy from K = 3. Cell pairs are matched on both initial capacity (±0.1 Ah) and early capacity slope (±0.02 Ah/diag) — ensuring they are indistinguishable by conventional metrics — then scored on whether $\tau_3$ slope or capacity slope correctly identifies which cell degrades further. $\tau_3$ slope achieves 71–82% accuracy across gap thresholds; capacity slope remains near chance (~58%). (e) Pareto front: prognostic MAE versus cumulative diagnostic time. LGN uses K × 40 s of pulse relaxation; capacity requires K × ~ 6 hr of 0.2C rate-performance testing. At K = 7, LGN achieves comparable MAE 540× faster. Severson et al. (2019) shown for reference (LFP chemistry, 124 cells, 100-cycle protocol). (f) Leave-one-cell-out parity at K = 7 (285 cells, ≥7 diagnostics). τ-only features: MAE = 9.1%, $\rho_s$ = 0.79; capacity features: MAE = 9.5%, $\rho_s$ = 0.82.

Within individual cells, time constants track degradation with high fidelity across this diverse fleet: the median within-cell Spearman $|\rho_s|$ is 0.976 for $\tau_2$ and 0.927 for $\tau_3$ across 320 cells with ≥5 diagnostics (Figure 2b). The critical question, however, is whether early time-constant trajectories predict end-of-life outcomes that capacity alone cannot. Figure 2a isolates this signal by matching cells on initial capacity (±0.15 Ah of the fleet median) to remove any starting-point advantage, then stratifying into terciles by $\tau_3$ slope computed from only the first three diagnostics. The separation is striking: cells in the lowest $\tau_3$-slope tercile reach 80% EOL within approximately 15 diagnostics, while cells in the highest tercile remain above 80% throughout. These cells started with indistinguishable capacity — the divergence is visible only in the time-constant trajectory. Figure 2d sharpens this claim with a controlled pairwise test. Cell pairs are matched on both initial capacity (±0.1 Ah) and early capacity slope (±0.02 Ah per diagnostic), ensuring they are indistinguishable by any conventional capacity-based metric. For each pair, we inquire if $\tau_3$ slope from K = 3 correctly identify which cell ultimately degrades further. Across gap thresholds from 10% to 40% ΔSOH at end-of-life, $\tau_3$ slope achieves 71–82% pairwise accuracy, while capacity slope — the only feature available to conventional methods on these matched pairs — hovers near chance at ~58%. This is direct evidence that time constants encode orthogonal prognostic information about degradation trajectory that capacity fundamentally cannot access from the same early-life window.

The practical consequence is a dramatic reduction in the diagnostic burden required for lifetime prediction (Figure 2e). LGN features derived from K pulses of 40-second relaxation are compared against capacity features derived from K rate-performance tests each requiring ~6 hours of 0.2C cycling, with both fed into identical ridge regression models via LOOCV. At K = 7, τ-only features (slopes and values of $\tau_1$, $\tau_2$, $\tau_3$ from the first 7 diagnostics; 6 features total) achieve 9.1% MAE — comparable to capacity-based prediction (9.5% MAE; 2 features value and slope) but from 4.7 minutes of cumulative pulse data versus 42 hours of rate testing, a 540× reduction in diagnostic time. Severson et al. (2019), shown for reference, achieved comparable accuracy on a smaller LFP fleet (124 cells) using 100 full charge–discharge cycles. The parity plot at K = 7 (Figure 2f) confirms that prediction quality is uniform across the SOH range: 285 cells spanning final outcomes from 30% to 95% SOH fall along the diagonal with no systematic bias toward well-behaved or deeply degraded cells, and τ-only features ($\rho_s$ = 0.79) match capacity features ($\rho_s$ = 0.82) despite requiring nearly three orders of magnitude less measurement time.

## Manufacturing quality control

Before a battery cell enters an EV pack or grid storage system, it passes through incoming inspection — a quality gate that must catch manufacturing defects without destroying the cell or consuming hours of test time. Current practice relies on capacity grading and DC resistance, measurements that are slow (full charge–discharge cycles) or insensitive to interfacial defects that only manifest later in life. We evaluate whether a single 40-second HPPC pulse, analysed by LGN, can serve as a rapid manufacturing fingerprint. The Samsung INR21700-50E dataset provides 254 commercially manufactured cells spanning 32 production batches, each measured once at 50% SOC and 25 °C across 8 tester channels [16].

LGN $\tau_1$ exhibits clear batch-level structure (Figure 3a). A k-means partition ($k = 2$) applied to batch-mean $\tau_1$ values identifies batches 21–25 as an anomalous subpopulation, with $\tau_1$ depressed ~21% relative to the normal-batch population mean ($\mu \approx 1.44$ s). This signal is not an instrumentation artifact: $\tau_1$ distributions are statistically indistinguishable across the 8 tester channels (ANOVA $F = 0.61$, $p = 0.74$; Figure 3b), ruling out systematic measurement bias as the source of the observed batch separation. The nature of the anomaly is physically specific. In anomalous batches, $\tau_1$ drops 21% while $\tau_2$ is essentially unaffected (−2%), localizing the defect to the fast interfacial process rather than bulk transport (Figure 3c). Pulse resistance $R_{pulse}$ rises 66% in the same batches — consistent with degraded surface kinetics — but the $\tau_1$ signal is not redundant with resistance. Across all 254 cells, $\tau_1$ and $R_{pulse}$ are strongly anti-correlated ($\rho = -0.75$, $p < 10^{-46}$; Fig. 3d), yet the anomalous batches deviate from the normal-population trend, indicating that $\tau_1$ = RC captures dynamic storage and transport information beyond R alone. This within-batch relationship holds in 31 of 32 batches, confirming it is not driven by the between-group separation. The most striking comparison is with conventional curve fitting. Kruskal–Wallis testing across all 32 batches yields H = 117 ($p < 10^{-11}$) for LGN $\tau_1$, providing overwhelming evidence of batch-level heterogeneity (Figure 3e). Curve-fit $\tau_1$ — extracted from the same pulse data by standard multi-exponential fitting — yields H = 35 (p = 0.28), failing to reject the null hypothesis that all batches are drawn from the same distribution. Both methods fit the pulse waveform; only LGN recovers a time constant stable and precise enough to resolve manufacturing variation at this scale.

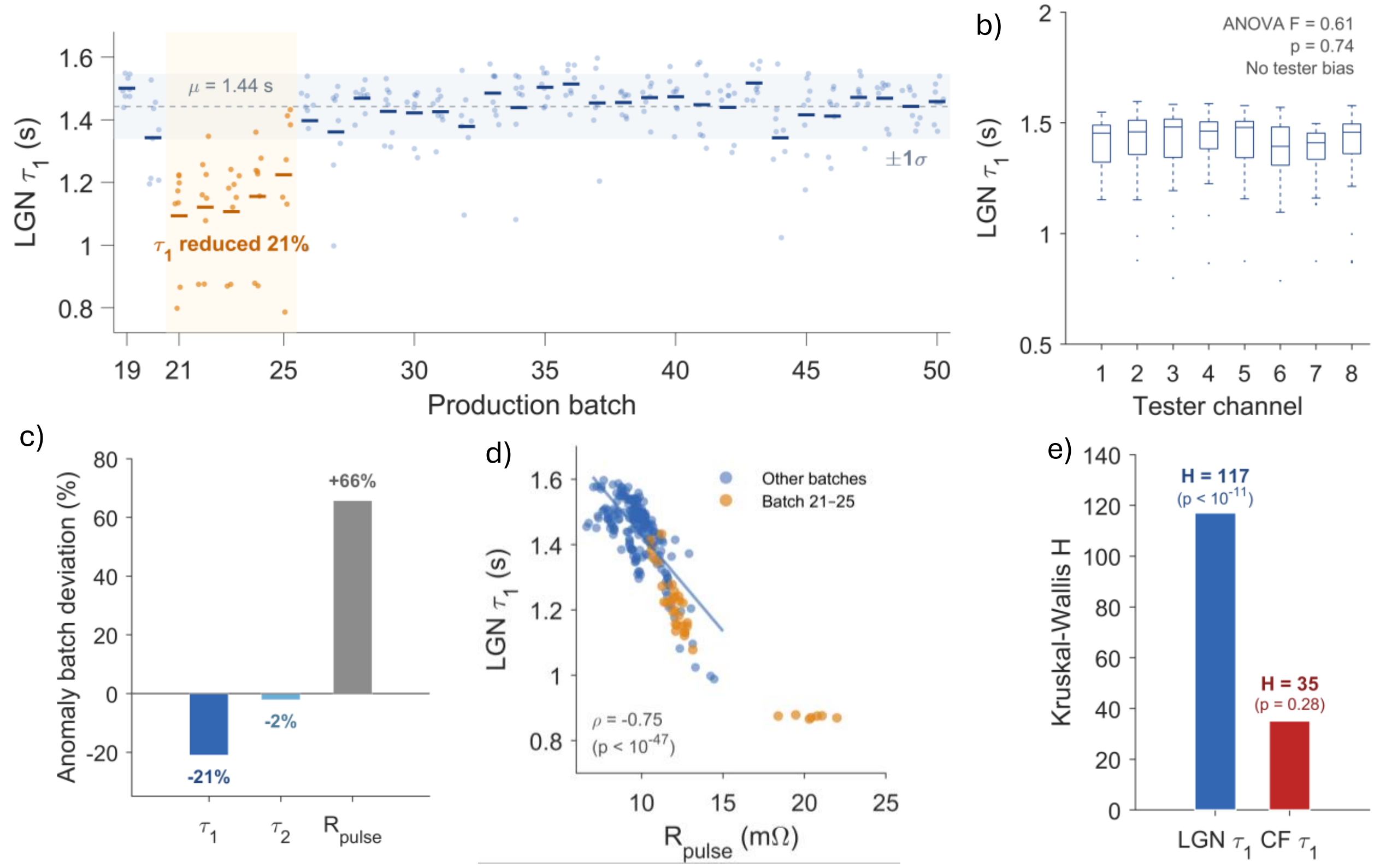

**Figure 3 | Manufacturing Quality Control: A single 40-second pulse recovers a manufacturing fingerprint at 254-cell scale (**Samsung INR21700-50E dataset, 32 production batches, 8 tester channels, one HPPC pulse per cell at 50% SOC, 25 °C) (a) LGN $\tau_1$ by production batch. Dashes: batch medians. Batches 21–25 identified by k-means ($k = 2$) on batch means; $\tau_1$ depressed 21% relative to normal batches. (b) Tester channel analysis. ANOVA confirms no instrument bias ($F = 0.61$, $p = 0.74$). (c) Mechanism separation. $\tau_1$ drops 21% in anomalous batches while $\tau_2$ is unaffected (−2%), localizing the defect to the fast interfacial process; $R_{pulse}$ rises 66%, consistent with degraded surface kinetics. (d) $\tau_1$ versus $R_{pulse}$ ($\rho = -0.75$, $p \sim 10^{-47}$); anomalous batches deviate from normal trend (e) Batch discrimination. Kruskal–Wallis $H = 117$ ($p < 10^{-11}$) for LGN $\tau_1$ versus $H = 35$ ($p = 0.28$) for curve-fit $\tau_1$; curve fitting cannot reject the null hypothesis that all batches are identical.

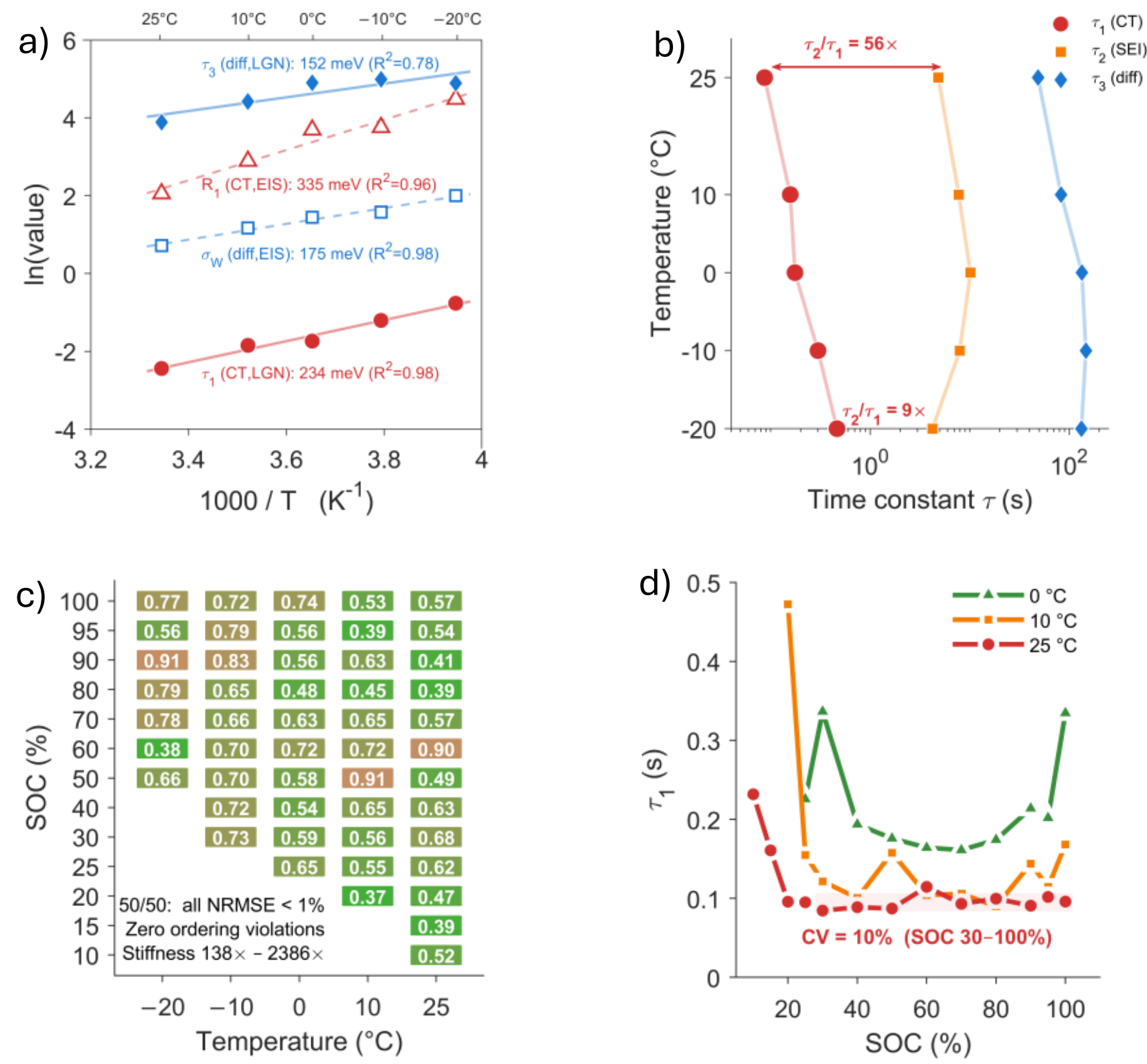

**Figure 4 | LGN recovers the electrochemical fingerprint of an NCA cell across 50 operating conditions with zero physics priors (**Panasonic NCR18650PF, 13 SOC levels, −20 °C to 25 °C). Each condition was analysed independently with a 3-state LGN model fit to a single HPPC pulse. (a) Cross-modal Arrhenius validation at SOC 50%. Filled markers: LGN ; open markers: EIS. The charge-transfer pair (red) yields Ea = 234 meV ($\tau_1$, LGN) and 335 meV ($R_1$, EIS); the diffusion pair (blue) yields Ea = 152 meV ($\tau_3$, LGN) and 175 meV ($\sigma_W$, EIS). (b) Eigenvalue spectrum at SOC 50%. The $\tau_2/\tau_1$ separation compresses from 56× at 25 °C to 9× at −20 °C as charge-transfer kinetics slow into the interfacial timescale. (c) Fit quality across all 50 conditions (NRMSE shown). Range: 0.37–0.91%, zero ordering violations ($\tau_1 < \tau_2 < \tau_3$), stiffness up to 2,386×. **(d)** $\tau_1$ vs SOC at three operating temperatures. At 25 °C, $\tau_1$ is effectively SOC-independent (CV = 10%, SOC 30–100%). At lower temperature, a characteristic U-shaped SOC dependence emerges, consistent with Butler-Volmer kinetics becoming concentration-limited as the Arrhenius factor suppresses the exchange current.

## Electrochemical validation

The results above demonstrate that LGN time constants are diagnostically and prognostically powerful — they track degradation, predict lifetime, and fingerprint manufacturing variation. But are they physically meaningful, or merely well-correlated statistical features? We address this by analysing a single Panasonic NCR18650PF NCA cell across 50 operating conditions spanning 13 SOC levels and five temperatures from −20 °C to 25 °C [17]. Each condition is fitted independently with a 3-state LGN model using a single HPPC pulse — no shared parameters, no physics priors, no knowledge of electrochemistry.

The most direct test of physical content is whether LGN time constants obey the Arrhenius equation, the universal signature of thermally activated processes. At SOC 50%, we pair each LGN quantity with its electrochemical counterpart measured independently by EIS (Figure 4a). For the charge-transfer process, LGN $\tau_1$ and EIS $R_1$ (from a 2RC fit to the high-frequency semicircle) both follow Arrhenius kinetics with $R^2 \geq 0.96$, but with different activation energies: $E_a$ = 234 meV for $\tau_1$ versus 335 meV for $R_1$. This discrepancy is not a failure of the method — it is a quantitative signature of the underlying physics. The charge-transfer resistance $R_1$ reflects the exchange current density through the Butler–Volmer equation, with an activation energy set by the desolvation barrier at the electrode–electrolyte interface. The time constant $\tau_1 = R_1C_1$ additionally encodes the temperature dependence of the double-layer capacitance $C_1$, which decreases at low temperature as reduced ionic mobility contracts the effective Helmholtz layer. Since $\ln(\tau) = \ln(R) + \ln(C)$, the Arrhenius slopes add: the 101 meV difference (335 − 234) isolates the capacitive thermal contribution — a quantity that EIS resistance alone cannot access. For solid-state diffusion, LGN $\tau_3$ ($E_a$ = 152 meV) and the EIS Warburg coefficient $\sigma_W$ ($E_a$ = 175 meV) agree more closely, consistent with diffusion being governed primarily by a single thermally activated hop rate with minimal capacitive modulation. That a method with zero electrochemical priors independently recovers activation energies consistent with decades of impedance literature — and does so from <60-second voltage traces rather than 30-minute frequency sweeps — is strong evidence that LGN is identifying genuine electrochemical processes rather than fitting artefacts.

The eigenvalue spectrum across temperature reveals the structural consequence of these differential activation energies (Figure 4b). At 25 °C, the three modes are well separated across three decades on the time-constant axis, with $\tau_2/\tau_1 = 56\times$. As temperature drops, charge transfer — the process with the highest activation energy — slows most dramatically, compressing the hierarchy until $\tau_2/\tau_1 = 9\times$ at −20 °C. This mode coalescence is observed most likely because at sufficiently low temperature, the fast electrochemical process is no longer fast. Despite this compression, LGN maintains sub-1% NRMSE across all 50 operating conditions with zero ordering violations ($\tau_1 < \tau_2 < \tau_3$ everywhere) and handles stiffness ratios up to 2,386× (Figure 4c). The eigenvalue ordering is never imposed and it emerges from the data at every condition, confirming that the three-mode decomposition reflects a genuine physical hierarchy rather than an arbitrary partitioning of the relaxation signal.

The SOC dependence of $\tau_1$ provides a final, independent consistency check (Figure 4d). At 25 °C, $\tau_1$ is effectively SOC-independent across the usable range (CV = 10%, SOC 30–100%), consistent with fast exchange kinetics that are not rate-limited by local lithium availability. At lower temperatures, a characteristic U-shaped SOC dependence emerges, with $\tau_1$ rising at both extremes of the SOC window. This is the expected signature of Butler–Volmer kinetics transitioning from activation-controlled to concentration-limited: as the Arrhenius factor suppresses the exchange current at low temperature, the charge-transfer rate becomes sensitive to the local surface concentration of lithium, which is depleted near 0% SOC and saturated near 100% SOC — both conditions that reduce the number of available reaction sites and slow the interfacial process. The temperature at which this transition becomes visible is itself informative: at 25 °C the thermal energy is sufficient to maintain activation control across the full SOC window, while at 0 °C the exchange current has dropped enough that concentration effects become rate-limiting at the extremes.

## Discussions

The central contribution of this work is not that pulse data can diagnose batteries but that the diagnostic information in pulse relaxation data and impedance spectroscopy are connected through a shared mathematical object: the eigenvalues of the relaxation generator. The same time constants that govern voltage decay in the time domain place poles in the impedance transfer function in the frequency domain. LGN makes this connection operational by providing a stable, training-free algorithm that extracts these eigenvalues from any pulse waveform, converting a laboratory-only measurement into something existing battery management hardware can perform in real time.

A real battery cell cannot be fully represented by three (or any small number of) discrete $RC$ pairs. Each electrode contains a distribution of particle sizes and tortuous transport pathways; the SEI/interphase is spatially heterogeneous; and electrolyte/solid diffusion introduces inherently distributed time scales. Each source of heterogeneity turns a single $RC$ process into a continuum of relaxation times. For example, diffusion time scales as particle radius squared, so even modest particle-size variation spreads the diffusion response across decades in $\tau$. Interphase thickness variation broadens interfacial/transport features similarly.

A standard representation is the distribution of relaxation times (DRT) $g(\tau)$, under which the post-pulse overpotential can be written as

$$\eta(t) = \int_0^{\infty} g(\tau)\ e^{-t/\tau}\, d\ln\, \tau,$$

where $g(\tau) \geq 0$ is a smooth spectral landscape with broad ridges rather than discrete spikes. In this setting, any finite-order model—$n$-RC ECM or $n$-state LGN—should be interpreted as a reduced-order approximation to a distributed system, positioned by the optimizer to capture the dominant spectral weight in the observation window. The objective function is formally identical to multi-exponential curve fitting; the distinction is structural. Curve fitting treats each time constant and amplitude as an independent scalar, producing an ill-conditioned landscape with flat directions that permit mode swapping, sign-indefinite amplitudes, and no stability guarantee. LGN constrains the problem through the generator: positivity is enforced by the softplus map, stability is guaranteed by construction, and gradients propagate through the matrix exponential rather than through individual basis functions, coupling all eigenvalues during optimization. The practical consequence is reproducibility — on the Popp dataset, this structural difference is the reason LGN resolves batch-level manufacturing variation that unconstrained fitting from identical data cannot (Figure 3e).

The theoretical framework (Supplementary Sections S1–S4) explains why this generalizes. The window scaling law shows that LGN eigenvalues are window-conditioned projections of the continuous relaxation spectrum. Their absolute values shift with observation window, but the degradation monotonicity argument explains why diagnostics are unaffected. Aging reshapes the entire spectral landscape coherently, so any fixed-window projection inherits a consistent trend with cell health (Supplementary S2, Table S1). Within-cell Spearman $|\rho_s|$ between $\tau_1$ and aging exceeds 0.91 at every window from 36 s to 3,600 s, confirming that the diagnostic ordering is preserved despite a >10× shift in absolute eigenvalue (Supplementary S3). The pole-locking mechanism clarifies when eigenvalues lock onto genuine spectral features rather than sliding along the continuum. Together, these explain why $\tau_1$ achieves $|\rho_s| = 0.999$ — not by identifying a single RC element, but by consistently sampling a spectral region that moves monotonically with degradation.

The τ-to-impedance correspondence is itself window-invariant: cross-validated Nyquist reconstruction achieves sub-1.3% MAPE at every window from 36 s to 3,600 s, and a single 36-second pulse reconstructs the full spectrum of an unseen cell at 1.0% error outperforming EIS 3RC self-reconstruction (2%) despite using no impedance data from the test cell (Fig S1e,f).

LGN requires a ≥1 Hz voltage trace during any rest period and a matrix exponential that executes in microseconds on embedded hardware. No potentiostat, training data, and chemistry-specific calibration required. This enables health monitoring at every rest event in deployed packs, single-pulse quality grading on production lines, rapid sorting of retired cells entering second-life markets, and early-warning prognosis before capacity fade becomes measurable. The diagnostic infrastructure is already deployed at terawatt-hour scale, the missing element was an algorithm that could read what the data already contains.

## Acknowledgements

This work was partially supported by the Department of Energy Grant No. DE-SC0022248, Office of Naval Research Grant No. N00014-21-1-2634, and Air Force Office of Scientific Research Grants No. FA9550-21-1-0305 and FA9550-22-1-0433.

## Supplementary:

### S1. Window Scaling Law

The fitted $\hat{\tau}_i$ are window-conditioned projections of the continuous spectrum $g(\tau)$, not intrinsic, window-invariant poles of the cell. The dependence of fitted eigenvalues on window length follows from a simple rescaling argument. Substitute $s = t/T$ so $dt = Tds$:

$$\int_0^T | \eta(t) - \sum_i v_{0,i}\, e^{-t/\tau_i} |^2 \, dt = T \int_0^1 | \eta(sT) - \sum_i v_{0,i}\, e^{-s/(\tau_i/T)} |^2 \, ds.$$

The prefactor $T$ does not affect the minimizer. Crucially, the exponentials depend on $\tau_i$ only through the dimensionless ratios

$$\theta_i \equiv \frac{\tau_i}{T}.$$

The optimizer does not see real time, it sees fractions of the observation horizon. When the underlying DRT is broad and smooth over the band probed by the window, the rescaled signal $\eta(sT)$ changes slowly with $T$, and the optimal $\hat{\theta}_i$ are approximately window-invariant. This yields the scaling law

$$\hat{\tau}_i \approx \hat{\theta}_i\, T \Rightarrow \hat{\tau}_i \propto T,$$

i.e., migration exponent $\alpha = d\log\, \hat{\tau}/d\log\, T \approx 1$ for modes tracking smooth spectral regions. Conversely, $\alpha \approx 0$ indicates modes locked onto narrow absolute-scale DRT peaks.

We test this on the Stanford SECL dataset [18] with LGN fitted at windows of 36 s, 360 s, and 3,600 s from the same pulse data (Figure S1b,d). The measured piecewise migration exponents (36→360 s, 360→3,600 s) are: $\tau_3$ (1.02, 1.01), $\tau_2$ (0.95, 0.60), $\tau_1$ (+0.80, −0.37). $\tau_3$ confirms the scaling law almost exactly: the median $\tau_3$ is 9.0 s at 36 s, 83.7 s at 360 s, and 1,008 s at 3,600 s — a 112× increase across a 100× window range. This is a pure DRT surrogate: the amplitude-weighted centroid of whatever slow relaxation content is visible in the window, not a fixed physical process. $\tau_2$ partially follows at short windows ($\alpha$ = 0.95) but decelerates at long windows ($\alpha$ = 0.60), suggesting it begins to resolve a genuine spectral feature as the observation horizon extends. $\tau_1$ tracks the scaling law from 36 to 360 s but reverses at long windows ($\alpha$ = −0.37) — the origin of this reversal is addressed in S3.

The physical mechanism underlying this scaling is energy weighting (Figure S1c). The $L^2$ energy contributed by a single mode:

$$E_T(c,\tau) = {v_0}^2 \int_0^T e^{-2t/\tau}\, dt = \frac{{v_0}^2\tau}{2}\left(1-e^{-2T/\tau}\right)$$

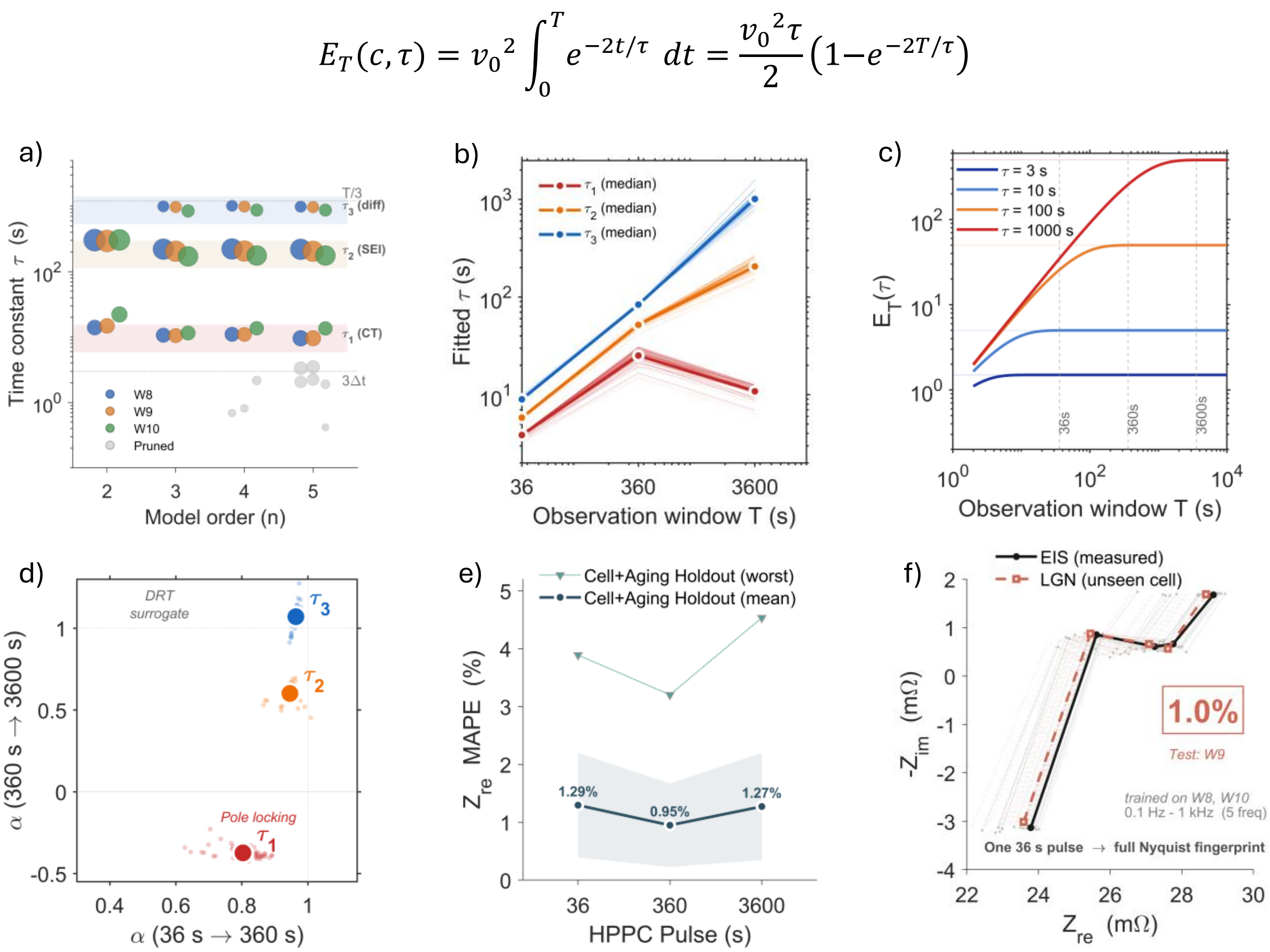

**Figure S1 | Window-conditioned LGN eigenvalue projection (Stanford SECL, 3 NMC cells × 15 aging states, SOC 50%).** (a) Three persistent modes emerge at n = 3; additional modes at n = 4, 5 are pruned or fall outside the identifiability window. (b) Eigenvalue migration across 36, 360, and 3,600 s windows. $\tau_3$ tracks slope = 1 (dashed); $\tau_1$ reverses at long windows. (c) Energy weighting: fast modes saturate, slow modes grow linearly with T, shifting optimizer attention. (d) Piecewise migration exponent α. $\tau_3$ confirms DRT surrogate ($\alpha \approx 1$); $\tau_1$ enters the pole-locking quadrant ($\alpha = -0.37$). (e) Nyquist reconstruction MAPE is window-invariant (1.29%, 0.95%, 1.27% under double-blind holdout). (f) Single 36 s pulse reconstructs the full Nyquist spectrum of an unseen cell at 1.0% MAPE.

For fast modes ($\tau \ll T$), $E_T \approx {v_0}^2\tau/2$ saturates. The mode has fully decayed and extending the window adds no information. For slow modes ($\tau \gg T$), $E_T \approx {v_0}^2 T$; the slow modes dominate the error budget with energy proportional to the window length. As T increases, the optimizer's attention shifts from fast to slow processes because that is where the reconstruction error concentrates.

**Table S1 | Within-window degradation tracking and cross-window rank consistency (Stanford SECL, 3 cells × 15 aging states, SOC 50%).** Left: mean within-cell Spearman $|\rho_s|$ between each eigenvalue and aging state. Right: mean within-cell Spearman $\rho_s$ between the same eigenvalue measured at different windows.

| Eigenvalue | 36 s | 360 s | 3600 s | 36↔360 | 360↔3600 | 36↔3600 |
|---|---|---|---|---|---|---|
| $\tau_1$ | 0.95 | 0.97 | 0.91 | 0.92 | 0.9 | 0.83 |
| $\tau_2$ | 0.72 | 0.68 | 0.94 | 0.76 | 0.67 | 0.60 |
| $\tau_3$ | 0.94 | 0.81 | 0.90 | 0.88 | -0.61 | -0.85 |

## S2. Degradation Monotonicity

Window dependence raises a concern: if $\hat{\tau}_i$ are not pure physical constants, how can they serve as reliable health indicators? The answer is that degradation does not move a single isolated peak—it reshapes the entire relaxation landscape. SEI growth shifts and broadens mid-frequency transport features. Loss of active surface area alters the charge-transfer peak. Lithium inventory depletion reshapes diffusion dynamics. These mechanisms are coupled: a thicker SEI changes the boundary conditions for charge transfer, which in turn modifies how lithium enters the particles. The result is a global spectral drift rather than a local perturbation.

Any finite-dimensional projection of a globally drifting spectrum inherits a consistent monotone trend. Whether the three fitted modes sit near 3–10 s (short window) or 10–1000 s (long window), they are sampling a spectrum that has moved coherently with aging. The absolute coordinates change with $T$; the ordering with aging does not. EIS is a high-resolution projection of the full landscape—valuable for mechanistic attribution, but often unnecessary if the question is simply whether the system has moved toward slower, more resistive dynamics.

Formally, let the DRT at aging state $k$ be $g_k(\tau)$, and suppose aging induces a monotone rightward shift and amplification on $\ln \tau$. Then for any fixed window $T$and any projected mode $i$, the extracted coordinate satisfies $\hat{\tau}_i(T; k+1) > \hat{\tau}_i(T; k)$(up to noise and re-binning), yielding strong within-window rank correlation with aging. Table S1 confirms this. Within-cell Spearman $|\rho_s|$ between $\tau_1$ and aging exceeds 0.91 at every window from 36 s to 3,600 s. Cross-window rank correlations for $\tau_1$ range from +0.83 to +0.92 — the diagnostic ordering is preserved despite a >10× shift in absolute eigenvalue. $\tau_3$ shows strong within-window tracking ($|\rho_s| \geq 0.81$) but inverted cross-window ranking ($\rho_s = -0.85$ between 36 s and 3,600 s), consistent with projections through opposite ends of the DRT: at short windows $\tau_3$ tracks the fast tail of slow processes, while at long windows it tracks the slow processes themselves. A BMS operating at a fixed pulse length will never see this inversion.

## S3. Re-binning and Pole Locking

$\tau_1$ breaks the window scaling law. Its piecewise migration exponents are (Figure S1d): 36 → 360 s: $\alpha = +0.80$ (surrogate behavior) 360 → 3,600 s: $\alpha = -0.37$ (reversal)

The median $\tau_1$ rises from 3.9 s to 25.2 s as the window extends from 36 to 360 s, then drops back to 10.8 s at 3,600 s. This is the signature of re-binning. With only three modes to approximate a continuum, the optimizer partitions the DRT into three bins. At 360 s, $\tau_2$ and $\tau_3$ cannot cover the full slow tail alone, so $\tau_1$ is drafted upward to reduce error. At 3,600 s, $\tau_2$ and $\tau_3$ have migrated to ~206 s and ~1,008 s respectively, absorbing enough of the slow continuum that $\tau_1$ is released. It snaps back to ~10.8 s and locks onto the narrow charge-transfer peak — a genuine spectral feature rather than a projection artifact.

The charge-transfer peak is the narrowest feature in the DRT because exchange current density depends on temperature, potential, and surface chemistry, but not on particle size. Unlike diffusion (where $\tau \propto r^2$ smears the peak across decades) or SEI transport (where thickness variation broadens the peak), charge transfer produces a relatively sharp spectral feature. When the optimizer has a degree of freedom to spare, it locks onto this narrow peak rather than sliding along the smooth continuum. A narrow peak acts like a discrete pole; a broad peak acts like a continuum that can only be approximated by a surrogate.

**S4. The $\tau$ = RC Amplification**

LGN directly extracts the time constants $\tau_i = R_i C_i$ (or their dynamical equivalents), whereas EIS often treats $R_i$and $C_i$ separately. This matters diagnostically because aging frequently changes both in reinforcing directions: interfacial resistances increase while effective capacitances (active surface area) decrease. The product $\tau = RC$ therefore amplifies co-evolving degradation signals into a single coordinate. This amplification is not a modeling trick; it is a physical consequence of the relaxation structure. Nyquist reconstruction accuracy provides a final, independent test of this diagnostic power of LGN. Under double-blind cell-plus-aging holdout — where neither the test cell nor its aging state appears in fitting (simple linear ridge regression; features: $\tau_1$ , $\tau_2$, $\tau_3$) — the mean MAPE is 1.29% at 36 s, 0.95% at 360 s, and 1.27% at 3,600 s (Figure 1e). The eigenvalues shift by two orders of magnitude, yet the impedance prediction error remains below 1.3% at every window. From a single 36-second pulse, LGN reconstructs the full Nyquist spectrum of an unseen cell (W9, trained on W8 and W10 only) at 1.0% MAPE across five frequencies spanning 0.1 Hz to 1 kHz — impedance information that conventionally requires a 30-minute frequency sweep. This is not merely competitive with EIS: LGN's cross-validated prediction on an unseen cell achieves lower MAPE than EIS 3RC self-reconstruction (1.2% vs 2%), where the model is fitted directly to each cell's own impedance data. The result is consistent with a learned linear map that captures population-level structure in the $\tau$-to-impedance relationship, effectively borrowing statistical strength across cells in a way that single-cell parametric fitting cannot. Finally, this generalization extends across SOC: cell-LOOCV reconstruction at SOC 20%, 50%, and 80% all yield MAPE below 1.2%, confirming that the $\tau$-to-impedance mapping is not SOC-specific but reflects an underlying electrochemical correspondence that holds across the operating envelope.